\documentclass[floatfix,lengthcheck,showpacs,amssymb,amsmath,amsfonts,twocolumn]{aastex63}
\usepackage{CJK}
\usepackage{lineno}
\usepackage[utf8]{inputenc}
\usepackage{color}
\usepackage{graphicx}
\usepackage{float}
\usepackage{subfigure}
\usepackage{amsmath}
\usepackage{afterpage}
\usepackage{acronym}
\usepackage{multirow}
\usepackage{tikz}
\usetikzlibrary{arrows,shapes,trees,decorations.pathreplacing}
\usepackage{import}
\usepackage{svn-multi}
\usepackage{lineno}
\usepackage{hyperref}
\usepackage{gensymb}
\hypersetup{
    colorlinks=true,
    linkcolor=blue,
    filecolor=magenta,      
    urlcolor=cyan,
}

\newcommand{\msun}{\ensuremath{\mathrm{M}_{\odot}}}

\begin{document}
\begin{CJK*}{UTF8}{gbsn}
\title[]{Search for gravitational waves from the coalescence of sub-solar mass and eccentric compact binaries}

\correspondingauthor{Alexander H. Nitz}
\email{alex.nitz@aei.mpg.de}

\author[0000-0002-1850-4587]{Alexander H. Nitz}
\author[0000-0002-2928-2916]{Yi-Fan Wang (王一帆)}
\affil{Max-Planck-Institut f{\"u}r Gravitationsphysik (Albert-Einstein-Institut), D-30167 Hannover, Germany}
\affil{Leibniz Universit{\"a}t Hannover, D-30167 Hannover, Germany}


\begin{abstract}
We present the first search for gravitational waves from sub-solar mass compact-binary mergers which allows for non-negligible orbital eccentricity. Sub-solar mass black holes are a signature of primordial origin black holes, which may be a component of dark matter. To produce binary coalescences, primordial black holes may form close binaries either in the early universe or more recently through dynamical interactions. A signature of dynamical formation would be the observation of non-circularized orbits. We search for black hole mergers where the primary mass is $0.1-7\msun$ and the secondary mass is $0.1-1\msun$. We allow for eccentricity up to $\sim0.3$ at a dominant-mode gravitational-wave frequency of 10 Hz for binaries with at least one component with mass $>0.5\msun$. We find no convincing candidates in the public LIGO data. The two most promising candidates have a false alarm rate of 1 per 3 and 4 years, respectively, which combined is only a $\sim 2.4\sigma$ deviation from the expected Poisson rate. Given the marginal statistical significance, we place upper limits on the rate of sub-solar mass mergers under the assumption of a null observation and compare how these limits may inform the possible dark matter contribution.
\end{abstract}

\section{Introduction}

\begin{figure}[tb!]
    \centering
    \includegraphics[width=\columnwidth]{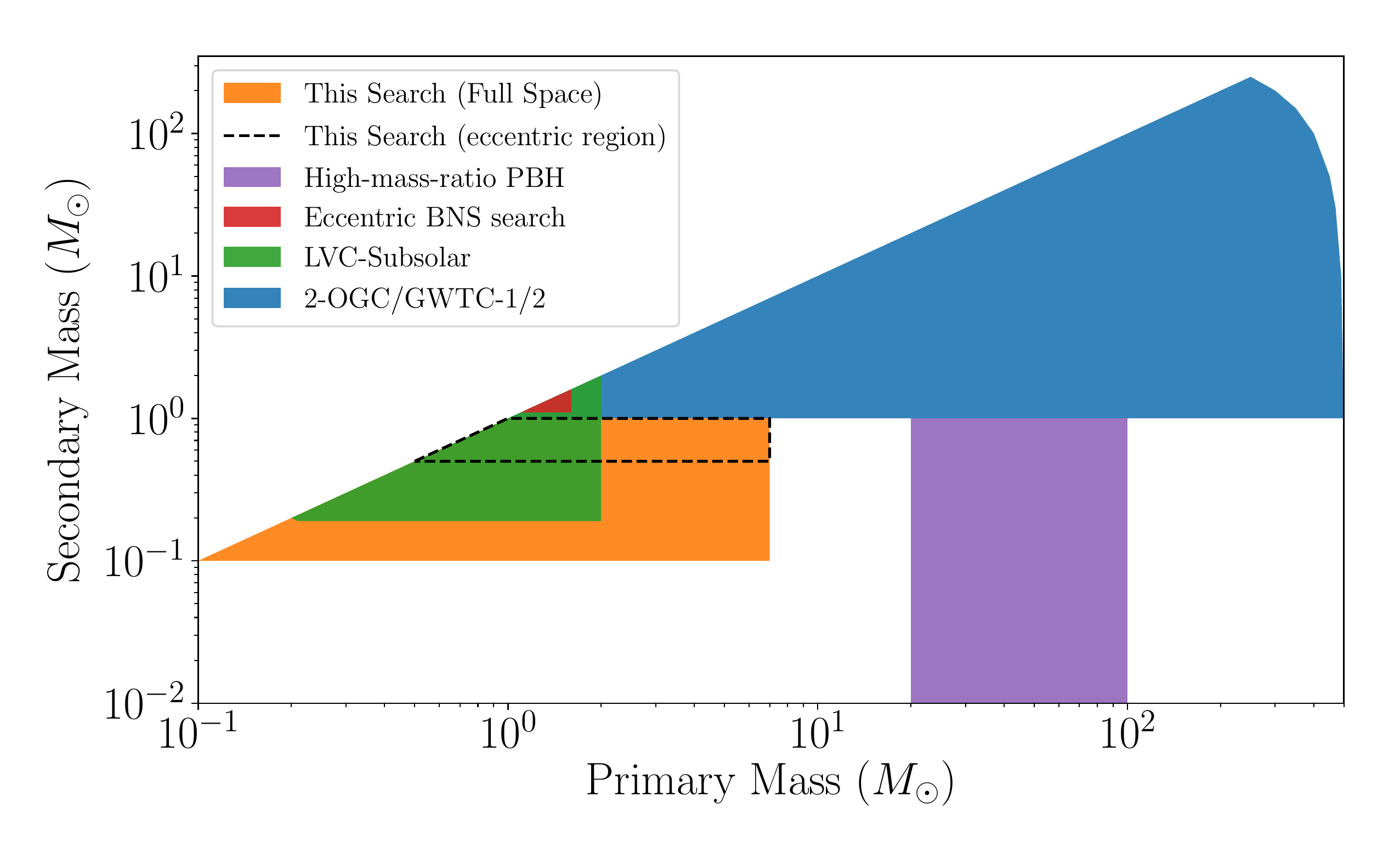}
    \caption{A comparison of gravitational-wave compact-binary searches using the LIGO and Virgo data. The extent of the detector-frame primary and secondary mass space is given for our full search (orange), the region we search for eccentric binaries (dotted lines), the high-mass-ratio sub-solar mass search  (purple) \citep{Nitz:2020bdb}, the binary neutron star eccentric search (red) \citep{Nitz:2019spj}, the LVC sub-solar mass search (green) \citep{Authors:2019qbw} and the standard stellar-mass searches such as 2-OGC/GWTC (blue) \citep{Nitz:2019hdf,GWTC2}. Note that several of these searches also allow for spin on the component-compact objects (not pictured). For our search, we allow for modest eccentric orbits ($e_{10}\lesssim0.3$) where at least one component has a mass $>0.5 \msun$.}
    \label{fig:searches}
\end{figure}

The field of gravitational-wave astronomy has been rapidly expanding ever since the first detection of gravitational waves with GW150914 in 2015 by the twin LIGO observatories in Hanford, WA and Livingston, LA~\citep{Abbott:2016blz}. In the last few years, there have been dozens of binary black hole mergers observed, two binary neutron stars mergers and possible black hole-neutron star mergers \citep{GWTC2}. There is a growing worldwide gravitational-wave network which in addition to the LIGO observatories~\citep{TheLIGOScientific:2014jea} has been joined by the Virgo observatory~\citep{TheVirgo:2014hva}. Shortly, the KAGRA observatory~\citep{KAGRA} will also begin joint observation~\citep{Aasi:2013wya}.

The high rate of black hole merger observations has sparked renewed interest in the possibility of primordial black holes contributing to dark matter~\citep{Bird:2016,Clesse:2016vqa,Sasaki:2016jop,Chen:2018czv,DeLuca:2020qqa}.  The growing population shows signs of an upper mass cutoff at $\sim40-50\msun$~\citep{GWTC2,Abbott:2020gyp,Roulet:2020wyq}, which would be consistent with pair-instability supernova~\citep{Woosley:2016hmi,Belczynski:2016jno,Marchant:2018kun,Woosley_2019,Stevenson:2019rcw}. However, one exceptional observation, GW190521, may lie within the resulting mass gap~\citep{Abbott:2020tfl} or straddle it~\citep{Fishbach:2020qag,Nitz:2020mga} and may show signs of an eccentric orbit~\citep{Gayathri:2020coq,Romero-Shaw:2020thy,CalderonBustillo:2020odh}. In addition, there is weak evidence from the overall population that suggests the observation of precession, in spite of the overall population being consistent with negligible effective spin~\citep{Abbott:2020gyp}. If confirmed, this would indicate that at least some fraction of the population has non-negligible spin which may not be preferred by many primordial black hole models~\citep{Chiba:2017rvs,DeLuca:2019buf,DeLuca:2020bjf,Mirbabayi:2019uph}. Both factors suggest that the observed binary black hole population would be difficult to explain with only primordial black holes, but would require a significant fraction of stellar-origin black holes. 

Given the speculative nature and uncertain mass distribution of primordial black holes~\citep{Jedamzik:1996mr,Widerin:1998my,Georg:2017mqk,Byrnes:2018clq}, it may be possible for a separate population of primordial black holes to exist, or for primordial black holes to contribute a fraction of the observed black hole mergers. However, distinguishing the standard stellar-origin population from a primordial population may be challenging in the case where their mass distributions overlap. To date, the majority of observed binary black hole mergers are near equal mass~\citep{GWTC2,GWTC2-rate}, with some notable exceptions ~\citep{Abbott:2020khf,LIGOScientific:2020stg}. No current observations have a component black hole with mass convincingly $<1\msun$. If a merger were found with at least one component definitely bounded below $\sim1\msun$, this would be a clear sign of primordial origin, as standard stellar evolution is not expected to produce such light black holes and neutron stars \citep{minmass1,minmass2}.

To produce gravitational-waves, primordial black holes must form a tight binary which will merge in time to be observed by the current observatories. Binaries may have formed in the early universe \citep{Nakamura:1997sm,Sasaki:2016jop}, in which case, their orbits would have circularized. However, binaries may also dynamically assemble through 2-body gravitational-wave braking~\citep{Bird:2016} or 3-body interactions~\citep{2012.03585} within dark matter halo structures in the late Universe. In these cases, there may be residual eccentricity which can be observed by ground-based detectors~\citep{Cholis:2016,Wang:2021qsu}. 

In this paper we search for gravitational waves from the coalescence of a sub-solar mass black hole $0.1-1 \msun$, which if observed would be most likely primordial in origin, with a black hole with mass $0.1-7\msun$. Past searches for primordial black holes by searching for mergers involving sub-solar mass black holes have so far yielded no detections~\citep{Abbott:2018oah,Authors:2019qbw,Nitz:2020bdb}. Compared to past analyses, we extend the search for comparable mass primordial black hole mergers down to $0.1\msun$, increase the upper limit on the primary mass, and for the first time, also search for primordial black holes mergers with residual eccentricity, which would be significant evidence for a dynamical capture formation channel. For computational reasons, we target sources with at least one component mass $>0.5\msun$ and orbital eccentricity $e<0.3$ at a fiducial dominant-mode gravitational-wave frequency of 10 Hz (orbital frequency of 5 Hz). A comparison of the region we search compared to past analyses is shown in Fig.~\ref{fig:searches}.

We find no statistically significant merger candidates and so place limits on the rate of mergers, finding that at $90\%$ confidence the rate of 0.1-0.1~\msun (1.0-1.0~\msun) mergers is $\lesssim1.7\times10^6$  ($5.5\times10^3$) Gpc$^{-3}$yr$^{-1}$ for circular binaries. The limit is matched for sources with moderate eccentricity $e_{10} < 0.3$ where our search has targeted.

Although highly model dependent~\citep{Nakamura:1997sm,Sasaki:2016jop,Bird:2016,10.1103/PhysRevD.99.043533,Clesse:2016vqa,Chen:2018czv,Ali-Haimoud:2017rtz}, limits on the merger rate can constrain the contribution fraction of primordial black holes to dark matter when assuming a particular formation mechanism. As an example, which may be compared to similar types of gravitational-wave searches, we find that if we assume the model of dynamical formation proposed in ~\cite{Chen:2018czv} for equal-mass binaries, we can place a 90\% limit on the contributing fraction of dark matter at $\leq 11\%(1\%)$ for $0.1(1)\msun$ sources. As current models predict only a small fraction of sources will have measurable eccentricity~\citep{Wang:2021qsu}, our null observation for moderately eccentric sources is consistent with the expectation.

\section{Search}

We search for gravitational waves using the currently available public LIGO data from the first and second observation runs~\citep{Vallisneri:2014vxa,Abbott:2019ebz}, which comprises $\sim164$ days of coincidence observing between the LIGO-Hanford and LIGO-Livingston observatories. We conduct the search using the open-source PyCBC-based archival analysis pipeline~\citep{Usman:2015kfa,pycbc-github}, which has been employed for similar analyses of both the public data~\citep{Nitz:2019hdf} and the analysis of proprietary data by the LIGO and Virgo collaborations~\citep{GWTC2}. Using similar configuration to both ~\citep{Nitz:2020bdb} and ~\citep{Nitz:2019hdf}, we use matched-filtering to extract a signal-to-noise time series from the data using models of the gravitational-wave signal, apply tests of signal-consistency and data quality, and finally identify and rank candidates. The statistical significance of any candidates is assessed using the standard method of creating symmetrically produced background analyses by applying non-astrophysical time offsets between the observing detectors~\citep{Usman:2015kfa}. 

Matched filtering is known to be an optimal method to extract signal from noise in the case of stationary Gaussian data and is the basis of the first stage of the search~\citep{findchirp}. This relies upon models of the expected gravitational-wave signal for any set of source parameters within the target search region. We model the gravitational waveform using the TaylorF2e model \citep{Moore:2019a,Moore:2019b,Moore:2018} which is an extension of the TaylorF2 model to include corrections for moderate eccentricity and models the inspiral, but not the merger or ringdown phase of the gravitational waveform. Given we search only up to a total mass of $8\msun$, the merger can be neglected as its corresponding gravitaitonal-wave frequency will be above the most sensitive band of the detectors $~20-500$Hz.

To search for a broad region, a discrete set of waveform templates is chosen using a stochastic algorithm~\citep{Harry:2009ea}, which ensures that at least $95\%$ of the SNR is recovered at any point in parameter space. The bank is sensitive to quasi-circular, nonspinning sources, with primary mass $0.1-7\msun$ and secondary mass $0.1-1\msun$. In addition, the bank is designed to search for non-circular, eccentric sources up to $e_{10}\sim0.3$  (defined at a reference gravitational-wave frequency of 10 Hz), where either black hole component has mass $>0.5\msun$. To control the computational cost, each template waveform is at most 512s in duration from its termination frequency down to a low frequency cutoff. The low frequency cutoff is chosen to enforce this duration limit with a minimum of 30 Hz. 

A region of our analysis was previously searched in ~\cite{Abbott:2018oah,Authors:2019qbw} by the LIGO and Virgo collaborations. Our search extends to a larger range of source masses and extends the analysis to search for eccentric binaries. For our common target region, we achieve minor sensitivity improvements by using a wider frequency band of the data (~\cite{Abbott:2018oah,Authors:2019qbw} searched using data only above 45 Hz). Notably, the analysis of ~\cite{Authors:2019qbw} also searched for sources with mild spin $\chi_\mathrm{eff} < 0.1$. We find that neglecting spin in our template bank is reasonable as primordial black holes are expected to have negligible spin~\citep{Chiba:2017rvs,DeLuca:2019buf,DeLuca:2020bjf,Mirbabayi:2019uph,Postnov:2019pkd}. In addition, nonspinning searches will still retain sensitivity to weakly spinning sources~\citep{Brown:2012gs,Nitz:2013mxa}.

\begin{table}
    \caption{The top five candidate events sorted by the false alarm rate of the search at each candidate's ranking statistic value. The chirp mass $\mathcal{M}$ of the candidate's associated template waveform is given in the detector frame. The SNR recovered by this template is reported for each LIGO detector. The most significant candidate is the previously identified BNS merger GW170817.}
    \label{table:candidates}
\begin{center}
\begin{tabular}{clrrrrrrrr}
GPS Time & FAR$^{-1}$ (y) & $\mathcal{M}$ & $e_{10}$  & $\rho_H$ & $\rho_L$ & \\ \hline
1187008882.45 & $>10^4$  & 1.20 & 0.08 & 17.6 & 23.9 \\ 
1172073536.21 & 4.2  & 0.50 & 0.26 & 6.0 & 7.6 \\
1127795466.48 & 3.5  & 0.50 & 0.00 & 7.1 & 5.9 \\
1135883588.06 & 0.7 & 0.41 & 0.00 & 7.1 & 6.5 \\
1134001579.85 & 0.4 & 0.61 & 0.22 & 6.4 & 6.5 \\
\end{tabular}
\end{center}
\end{table}

\section{Observational Results}

\begin{figure}[tb!]
    \centering
    \includegraphics[width=\columnwidth]{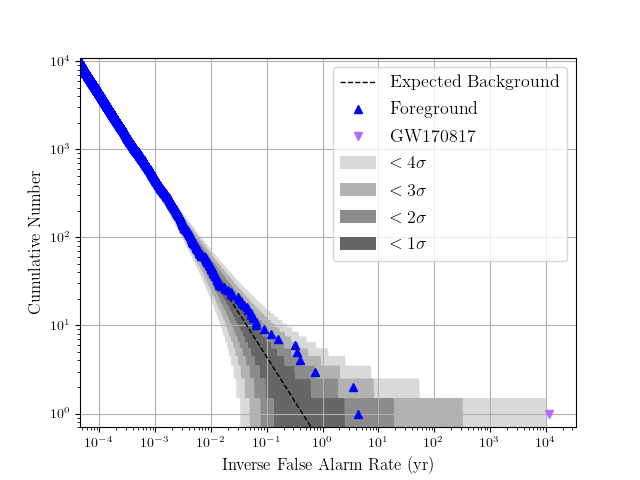}
    \caption{The cumulative number of candidates as a function of the inverse false alarm rate. The top candidate in our search was the previously detected GW170817 binary neutron star merger (purple), which has been excluded from the remainder of the candidate foreground. The two most significant following candidates represent only a $~2\sigma$ statistical deviation so we consider these results consistent with a null observation.}
    \label{fig:ifar}
\end{figure}

We search the public LIGO data~\citep{Vallisneri:2014vxa,Abbott:2019ebz} for sub-solar mass primordial black hole candidates. This includes just over 164 days of coincident observing by the LIGO-Hanford and LIGO-Livingston observatories. The top several candidates are given in Table~\ref{table:candidates} and the cumulative distribution of observed events is shown in Fig.~\ref{fig:ifar}.

The most significant candidate is the previously identified binary neutron star merger GW170817~\citep{TheLIGOScientific:2017qsa}. Our search is sensitive to this source as the mass ratio is poorly constrained~\citep{Abbott:2018wiz} for light sources where the merger frequency is not directly observed. Given the clear prior identification as a binary neutron star merger, we exclude this event from further analysis. 

The next two most significant candidates are identified at a false alarm rate of 1 per 4 and 1 per 3 years, respectively. Individually, either one would have a statistical significance of only $p\sim0.1-0.2$. If we consider the probability of the population excursion assuming a Poisson rate of candidates under the null hypothesis, we find that even in this case they represent only $\sim 2.4\sigma$ deviation. Given the statistical significance and strong prior odds one should place against the observation of a primordial black hole merger, we find our results are consistent with a null observation. If these were astrophysical in origin, this would imply a high rate of mergers which should be observed at high confidence in future observing runs. 

 \begin{figure}[tb!]
    \centering
    \includegraphics[width=\columnwidth]{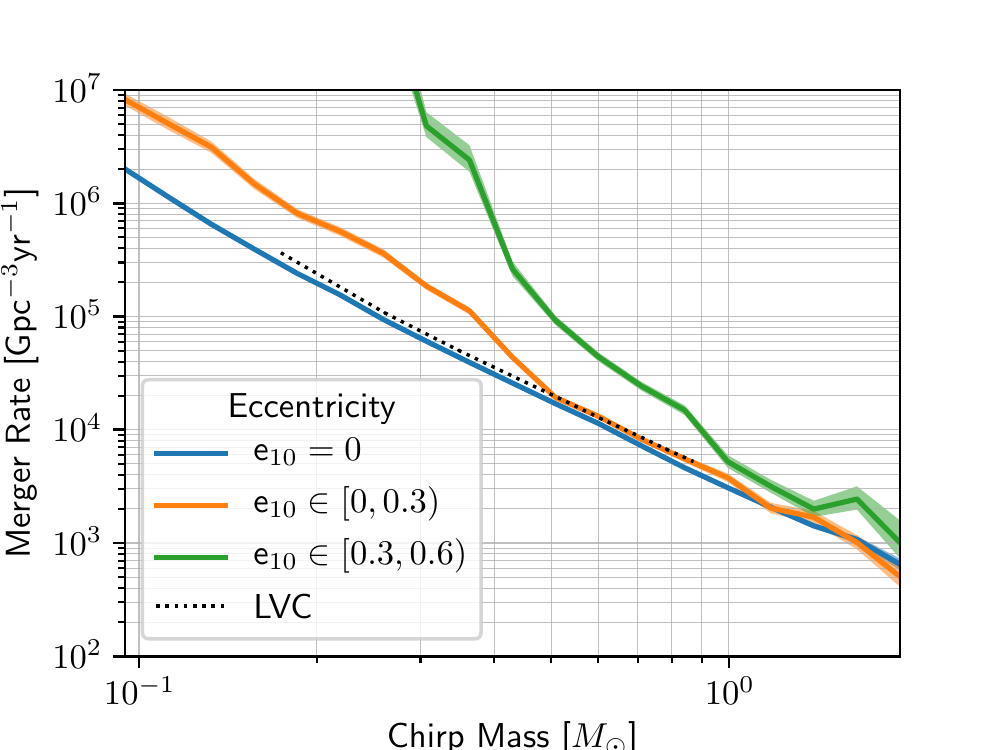}
    \caption{The upper limit on the merger rate for sources at 90\% confidence as a function of their chirp mass. The rate is shown for different fiducial eccentricities $e_{10}$ which are either zero (blue), uniform from 0-0.3 (orange) or uniform from 0.3-0.6. The effect of accounting for eccentricity in the search for sources with component mass above 0.5\msun~is evident in the orange curve as it begins to follow the non-eccentric limits above this threshold. The comparable LVC limits on non-eccentric mergers are shown with a black, dotted line.}
    \label{fig:rate}
\end{figure}

\section{Merger Rate}

Using our null detection, we place limits on the rate of binary mergers. We assign an upper limit at $90\%$ confidence using the loudest event method~\citep{Biswas:2007ni}, whereby the limit $R_{90}$ is given as
\begin{equation}
    R_{90} = \frac{2.3}{VT}
\end{equation}
where V is the estimated sensitive volume of the analysis to a chosen source population assessed at the false alarm rate of the most significant observed candidate, and $T$ is the length of the observation period. To measure the sensitivity of our analysis, we search for a simulated population of $O(10^5)$ sources.

The upper limit on the merger rate as a function of chirp mass is shown in Fig.~\ref{fig:rate}. Similar to the conclusions of~\cite{Authors:2019qbw} we find that this limit also holds for both equal-mass and non-equal mass sources with the same chirp mass, assuming they lie within the overall search region. For equal mass, non-eccentric sources, our results are consistent with the previous search conducted by the LVC~\citep{Authors:2019qbw}.

\section{Implications for primordial black hole abundance}

Our null search results can be used to place constraints on the fraction of dark matter composed of primordial black holes. 
This requires the use of a specific astrophysical model which predicts the merger rate from the initial abundance and distribution. 
Existing models widely vary in these predictions and have significant modelling uncertainties~\citep{Nakamura:1997sm,Sasaki:2016jop,Bird:2016,10.1103/PhysRevD.99.043533,Clesse:2016vqa,Chen:2018czv,Ali-Haimoud:2017rtz}. Eccentric binaries may be produced with a late time, dynamical formation scenario such as examined in~\cite{Wang:2021qsu}, however a null observation is consistent with current estimates of the detection rate, and so current observations do not yet constrain the primordial black hole contribution from this channel.

To provide an example which compares our results to similar limits in the gravitational-wave literature, we consider the mechanism proposed by \cite{Nakamura:1997sm,Sasaki:2016jop} where binary primordial black holes form in the early Universe and merge recently. The same model was used by \cite{Abbott:2018oah,Authors:2019qbw} to constrain the primordial black hole fraction for equal-mass binaries and \cite{Nitz:2020bdb} for high-mass-ratio binaries.
Note that in this scenario all the binaries would have been circularized to be detected in the local Universe. The binary merger rate for a general mass density distribution $P(m)$ is given by \cite{Chen:2018czv,Ali-Haimoud:2017rtz} as
\begin{eqnarray}\label{eq:pbhrate}
&R&(f_\mathrm{PBH},m_1,m_2) = 3\times10^6 f_\mathrm{PBH}^2(0.7f_\mathrm{PBH}^2 + \sigma_\mathrm{eq}^2)^{-\frac{21}{74}} \nonumber \\
&\times&(m_1m_2)^{\frac{3}{37}}(m_1+m_2)^{\frac{36}{37}}\mathrm{min}\left(\frac{P(m_1)}{m_1},\frac{P(m_2)}{m_2}\right
) \nonumber \\
&\times&\left(\frac{P(m_1)}{m_1}+\frac{P(m_2)}{m_2}\right) ~\mathrm{Gpc}^{-3} \mathrm{yr}^{-1}
\end{eqnarray}
where binary component masses $m_{1/2}$ are in units of $\msun{}$.
The parameter $\sigma_{eq}$ is the variance of dark matter density perturbation at the matter radiation equality epoch and takes the value $0.005$ according to \cite{Ali-Haimoud:2017rtz}. 

 \begin{figure}[tb!]
    \centering
    \includegraphics[width=\columnwidth]{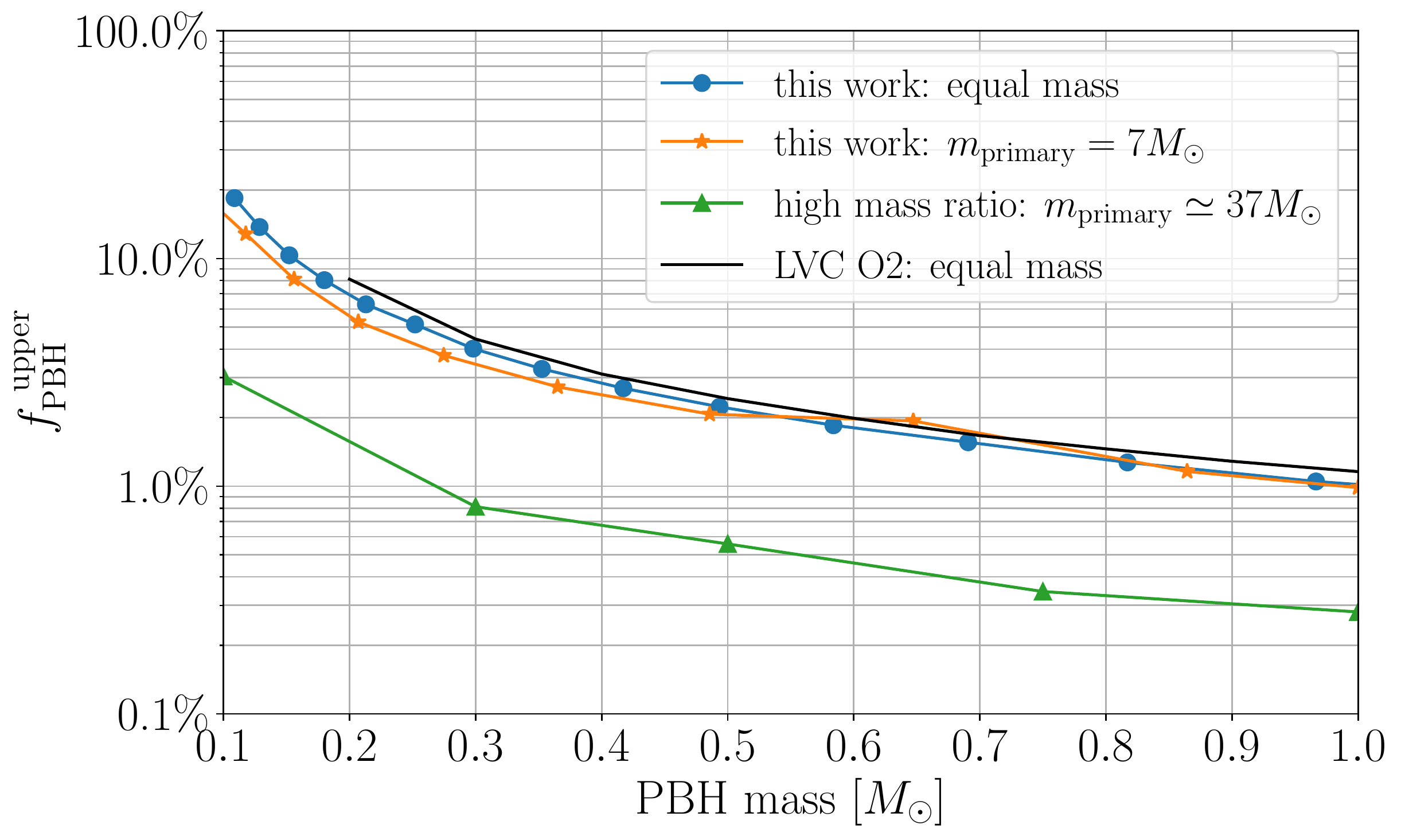}
    \caption{The upper limits on the fraction of primordial black hole in dark matter for the merger rate model proposed by \cite{Nakamura:1997sm,Sasaki:2016jop}.
    We consider both equal mass binaries and unequal mass binaries where the primary mass is fixed to $7\msun$.
    As a comparison, we also plot the constraints from \cite{Authors:2019fue} for equal mass binaries and \cite{Nitz:2020bdb} for high mass ratio binaries where the primary mass is fixed to the average of the 2-OGC events $\simeq 37\msun$.
    }
    \label{fig:pbhfraction}
\end{figure}

We first consider the assumption of single mass for primordial black holes, which is a fiducial assumption used by a variety of astrophysical constraints \citep{Green:2020jor} on primordial black hole abundance. 
In this case, the local merger rate in $Eq.(\ref{eq:pbhrate})$ reduces to $R(f_\mathrm{PBH},m) =3\times10^6 f_\mathrm{PBH}^2(0.7f_\mathrm{PBH}^2 + \sigma_\mathrm{eq}^2)^{-\frac{21}{74}}m^{-32/37}$.
In sub-solar mass region $[0.1,1] \msun$, applying the condition $R(f_\mathrm{PBH}^\mathrm{upper},m) =  R_{90}$ where $R_{90}$ is for $e_{10}=0$, the upper limit on the fraction of equal mass primordial black hole binaries is shown in Fig.\ref{fig:pbhfraction}.
For comparison, Fig.\ref{fig:pbhfraction} also plots the constraint extracted from the advanced LIGO/Virgo O2 sub-solar mass search results \citep{Authors:2019qbw}, but note this result is slightly tighter than the original one by using the more recent formula $Eq.(\ref{eq:pbhrate})$ which, compared with the event rate model in \cite{Sasaki:2016jop}, additionally includes the effect 
of torques on a binary from all other primordial black holes and linear density perturbations of the early Universe.

To investigate the effects of different mass distributions, we consider the alternate assumption that primordial black holes have two classes of mass and focus on the merger of unequal mass binaries. We consider the case where the primary mass is fixed to $7 \msun$ and the secondary mass is a fixed value which we allow to vary from $[0.1,1] \msun$.
To constrain $f_\mathrm{PBH}$, the ratio of abundance between two classes of mass needs to be specified.
We determine it by requiring 
\begin{equation}\label{eq:twomasscon}
    R\Big(f_\mathrm{PBH}^\mathrm{primary},m=7\msun\Big)= 12~ \mathrm{Gpc}^{-3}\mathrm{yr}^{-1},
\end{equation} 
where the left hand side is the equal-mass binary merger rate and the right hand side is obtained by considering the rate limit from GWTC-2 \citep{GWTC2}. This is the most optimistic rate which is consistent with either a small number of observations or the non-detection of the near equal-mass mergers of primordial black holes with $7\msun$.
Eq.~(\ref{eq:twomasscon}) results in $f_\mathrm{PBH}^\mathrm{primary} = 0.1\%$. 

With the condition from Eq.~\ref{eq:twomasscon}, the constraint on the fraction of sub-solar mass primordial black holes $f_\mathrm{PBH}^\mathrm{secondary}$ is plotted in Fig.~\ref{fig:pbhfraction}.
The previous constraints from the~\cite{Nitz:2020bdb} search are also plotted as a comparison, where the primary mass is fixed to be the average of the detections ($\simeq 37\msun$) in the 2-OGC catalog \citep{Nitz:2019hdf}, under the assumption that a majority of previous LIGO-observed black holes are primordial in origin, which would set the abundance of a portion of the mass function. This corresponds to $f_\mathrm{PBH}^\mathrm{primary} = 0.33\%$.
Results in Fig.~\ref{fig:pbhfraction} show that our constraints on equal mass primordial black holes binaries are slightly tighter than \cite{Authors:2019qbw} due to our increased volume sensitivity.
Constraints from unequal mass binaries with primary mass $7\msun$ are comparable to the equal mass case.
Upper limits from high mass ratio binaries are still the most stringent among direct searches for gravitational-waves from mergers due to the louder signals.

\section{Conclusions}

We conducted a search for gravitational-waves from sub-solar mass compact-binary mergers which allows for non-negligible orbital eccentricity. We found no promising candidates, and thus placed improved upper limits on the merger rate of compact-object binaries and the inferred abundance of primordial black holes. We relate our rate constraints to the abundance of primordial black hole by considering a specific astrophysical model predicting the merger rate. 
Our non-detection is consistent with the current tightest constraints on the primordial black hole abundance by previous gravitational wave direct searches~\citep{Abbott:2018oah,Authors:2019qbw,Nitz:2020bdb}.

Primordial black hole binaries which form during the early Universe and merge recently would have fully circularized, whereas the binaries formed in the late Universe by two body dynamical capture may be able to retain non-zero eccentricity due to quick merger after binary formation. However, as investigated in \cite{Wang:2021qsu}, the merger rate of this late Universe scenario ($\mathcal{O}(10^2)$ Gpc$^{-3}$ year $^{-1}$) is a few orders of magnitude lower than the $R_{90}$ we constrained for eccentric binary coalescence. Thus we conclude that our search results are consistent with this binary formation scenario. If any sub-solar mass candidates with eccentricity had been identified this would suggest a recent dynamical formation which could not be accounted for by this simple model.

Advanced LIGO and Virgo are continually being upgraded~\citep{livingreviewligo} and the third generation of gravitational-wave detectors are expected to further improve the sensitive volume by $\sim10^3$~\citep{et,cewhitepaper}. 
We expect the constraints on sub-solar mass binary black hole abundance to be $~10^{3-4}$ times tighter than the current search, assuming a null result. 
Moreover, if third generation detectors can achieve their challenging low frequency sensitivity targets, they will push the sensitive band down to $\sim 2-5$ Hz, where less eccentricity has been lost due to gravitational radiation.

\acknowledgments

 We acknowledge the Max Planck Gesellschaft. We thank the computing team from AEI Hannover for their significant technical support. This research has made use of data from the Gravitational Wave Open Science Center (https://www.gw-openscience.org), a service of LIGO Laboratory, the LIGO Scientific Collaboration and the Virgo Collaboration. LIGO is funded by the U.S. National Science Foundation. Virgo is funded by the French Centre National de Recherche Scientifique (CNRS), the Italian Istituto Nazionale della Fisica Nucleare (INFN) and the Dutch Nikhef, with contributions by Polish and Hungarian institutes. 
 
The configuration files and template bank necessary to reproduce the search are available at \url{https://github.com/gwastro/subsolar-ecc-primordial-search}
 
\bibliography{references}

\begin{thebibliography}{}
\expandafter\ifx\csname natexlab\endcsname\relax\def\natexlab#1{#1}\fi
\providecommand{\url}[1]{\href{#1}{#1}}
\providecommand{\dodoi}[1]{doi:~\href{http://doi.org/#1}{\nolinkurl{#1}}}
\providecommand{\doeprint}[1]{\href{http://ascl.net/#1}{\nolinkurl{http://ascl.net/#1}}}
\providecommand{\doarXiv}[1]{\href{https://arxiv.org/abs/#1}{\nolinkurl{https://arxiv.org/abs/#1}}}

\bibitem[{Aasi {et~al.}(2015)}]{TheLIGOScientific:2014jea}
Aasi, J., {et~al.} 2015, Class. Quantum Grav., 32, 074001,
  \dodoi{10.1088/0264-9381/32/7/074001}

\bibitem[{Abbott {et~al.}(2017)}]{TheLIGOScientific:2017qsa}
Abbott, B., {et~al.} 2017, Phys. Rev. Lett., 119, 161101,
  \dodoi{10.1103/PhysRevLett.119.161101}

\bibitem[{Abbott {et~al.}(2018{\natexlab{a}})}]{Abbott:2018oah}
---. 2018{\natexlab{a}}, Phys. Rev. Lett., 121, 231103,
  \dodoi{10.1103/PhysRevLett.121.231103}

\bibitem[{Abbott {et~al.}(2019{\natexlab{a}})}]{Authors:2019qbw}
---. 2019{\natexlab{a}}, Phys. Rev. Lett., 123, 161102,
  \dodoi{10.1103/PhysRevLett.123.161102}

\bibitem[{Abbott {et~al.}(2019{\natexlab{b}})}]{Authors:2019fue}
---. 2019{\natexlab{b}}, Astrophys. J., 886, 75,
  \dodoi{10.3847/1538-4357/ab4b48}

\bibitem[{Abbott {et~al.}(2016{\natexlab{a}})}]{Abbott:2016blz}
Abbott, B.~P., {et~al.} 2016{\natexlab{a}}, Phys. Rev. Lett., 116, 061102,
  \dodoi{10.1103/PhysRevLett.116.061102}

\bibitem[{Abbott {et~al.}(2016{\natexlab{b}})}]{Aasi:2013wya}
---. 2016{\natexlab{b}}, Living Rev. Relat., 19, 1, \dodoi{10.1007/lrr-2016-1}

\bibitem[{Abbott {et~al.}(2018{\natexlab{b}})}]{livingreviewligo}
---. 2018{\natexlab{b}}, Living Rev. Rel., 21, 3,
  \dodoi{10.1007/s41114-018-0012-9, 10.1007/lrr-2016-1}

\bibitem[{Abbott {et~al.}(2019{\natexlab{c}})}]{Abbott:2018wiz}
---. 2019{\natexlab{c}}, Phys. Rev. X, 9, 011001,
  \dodoi{10.1103/PhysRevX.9.011001}

\bibitem[{Abbott {et~al.}(2020{\natexlab{a}})}]{GWTC2}
Abbott, R., {et~al.} 2020{\natexlab{a}}.
\newblock \doarXiv{2010.14527}

\bibitem[{Abbott {et~al.}(2020{\natexlab{b}})}]{Abbott:2020gyp}
---. 2020{\natexlab{b}}.
\newblock \doarXiv{2010.14533}

\bibitem[{Abbott {et~al.}(2020{\natexlab{c}})}]{Abbott:2020tfl}
---. 2020{\natexlab{c}}, Phys. Rev. Lett., 125, 101102,
  \dodoi{10.1103/PhysRevLett.125.101102}

\bibitem[{Abbott {et~al.}(2020{\natexlab{d}})}]{GWTC2-rate}
---. 2020{\natexlab{d}}.
\newblock \doarXiv{2010.14533}

\bibitem[{Abbott {et~al.}(2020{\natexlab{e}})}]{Abbott:2020khf}
---. 2020{\natexlab{e}}, Astrophys. J. Lett., 896, L44,
  \dodoi{10.3847/2041-8213/ab960f}

\bibitem[{Abbott {et~al.}(2020{\natexlab{f}})}]{LIGOScientific:2020stg}
---. 2020{\natexlab{f}}, Phys. Rev. D, 102, 043015,
  \dodoi{10.1103/PhysRevD.102.043015}

\bibitem[{Abbott {et~al.}(2021)Abbott, Abbott, Abraham, Acernese, Ackley,
  Adams, Adhikari, Adya, Affeldt, Agathos, Agatsuma, Aggarwal, Aguiar, Aich,
  Aiello, Ain, Ajith, Allen, Allocca, Altin, Amato, Anand, Ananyeva, Anderson,
  Anderson, Angelova, Ansoldi, Antier, Appert, Arai, Araya, Areeda, Arène,
  Arnaud, Aronson, Arun, Ascenzi, Ashton, Aston, Astone, Aubin, Aufmuth,
  AultONeal, Austin, Avendano, Babak, Bacon, Badaracco, Bader, Bae, Baer,
  Baird, Baldaccini, Ballardin, Ballmer, Bals, Balsamo, Baltus, Banagiri,
  Bankar, Bankar, Barayoga, Barbieri, Barish, Barker, Barkett, Barneo, Barone,
  Barr, Barsotti, Barsuglia, Barta, Bartlett, Bartos, Bassiri, Basti, Bawaj,
  Bayley, Bazzan, Bécsy, Bejger, Belahcene, Bell, Beniwal, Benjamin, Bentley,
  Bergamin, Berger, Bergmann, Bernuzzi, Berry, Bersanetti, Bertolini,
  Betzwieser, Bhandare, Bhandari, Bidler, Biggs, Bilenko, Billingsley, Birney,
  Birnholtz, Biscans, Bischi, Biscoveanu, Bisht, Bissenbayeva, Bitossi,
  Bizouard, Blackburn, Blackman, Blair, Blair, Blair, Bobba, Bode, Boer,
  Boetzel, Bogaert, Bondu, Bonilla, Bonnand, Booker, Boom, Bork, Boschi, Bose,
  Bossilkov, Bosveld, Bouffanais, Bozzi, Bradaschia, Brady, Bramley, Branchesi,
  Brau, Breschi, Briant, Briggs, Brighenti, Brillet, Brinkmann, Brockill,
  Brooks, Brooks, Brown, Brunett, Bruno, Bruntz, Buikema, Bulik, Bulten,
  Buonanno, Buskulic, Byer, Cabero, Cadonati, Cagnoli, Cahillane, Bustillo,
  Callaghan, Callister, Calloni, Camp, Canepa, Cannon, Cao, Cao, Carapella,
  Carbognani, Caride, Carney, Carullo, Diaz, Casentini, Castañeda, Caudill,
  Cavaglià, Cavalier, Cavalieri, Cella, Cerdá-Durán, Cesarini, Chaibi,
  Chakravarti, Chan, Chan, Chao, Charlton, Chase, Chassande-Mottin, Chatterjee,
  Chaturvedi, Chen, Chen, Chen, Cheng, Cheong, Chia, Chiadini, Chierici,
  Chincarini, Chiummo, Cho, Cho, Cho, Christensen, Chu, Chua, Chung, Chung,
  Ciani, Ciecielag, Cieślar, Ciobanu, Ciolfi, Cipriano, Cirone, Clara, Clark,
  Clearwater, Clesse, Cleva, Coccia, Cohadon, Cohen, Colleoni, Collette,
  Collins, Colpi, Constancio, Conti, Cooper, Corban, Corbitt, Cordero-Carrión,
  Corezzi, Corley, Cornish, Corre, Corsi, Cortese, Costa, Cotesta, Coughlin,
  Coughlin, Coulon, Countryman, Couvares, Covas, Coward, Cowart, Coyne, Coyne,
  Creighton, Creighton, Cripe, Croquette, Crowder, Cudell, Cullen, Cumming,
  Cummings, Cunningham, Cuoco, Curylo, Canton, Dálya, Dana,
  Daneshgaran-Bajastani, D’Angelo, Danilishin, D’Antonio, Danzmann,
  Darsow-Fromm, Dasgupta, Datrier, Dattilo, Dave, Davier, Davies, Davis, Daw,
  DeBra, Deenadayalan, Degallaix, {De Laurentis}, Deléglise, Delfavero, {De
  Lillo}, {Del Pozzo}, DeMarchi, D’Emilio, Demos, Dent, {De Pietri}, {De
  Rosa}, {De Rossi}, DeSalvo, {de Varona}, Dhurandhar, Díaz, Diaz-Ortiz,
  Dietrich, {Di Fiore}, {Di Fronzo}, {Di Giorgio}, {Di Giovanni}, {Di
  Giovanni}, {Di Girolamo}, {Di Lieto}, Ding, {Di Pace}, {Di Palma}, {Di
  Renzo}, Divakarla, Dmitriev, Doctor, Donovan, Dooley, Doravari, Dorrington,
  Downes, Drago, Driggers, Du, Ducoin, Dupej, Durante, D’Urso, Dwyer, Easter,
  Eddolls, Edelman, Edo, Edy, Effler, Ehrens, Eichholz, Eikenberry, Eisenmann,
  Eisenstein, Ejlli, Errico, Essick, Estelles, Estevez, Etienne, Etzel, Evans,
  Evans, Ewing, Fafone, Fairhurst, Fan, Farinon, Farr, Farr, Fauchon-Jones,
  Favata, Fays, Fazio, Feicht, Fejer, Feng, Fenyvesi, Ferguson,
  Fernandez-Galiana, Ferrante, Ferreira, Ferreira, Fidecaro, Fiori, Fiorucci,
  Fishbach, Fisher, Fittipaldi, Fitz-Axen, Fiumara, Flaminio, Floden, Flynn,
  Fong, Font, Forsyth, Fournier, Frasca, Frasconi, Frei, Freise, Frey, Frey,
  Fritschel, Frolov, Fronzè, Fulda, Fyffe, Gabbard, Gadre, Gaebel, Gair,
  Galaudage, Ganapathy, Gaonkar, García-Quirós, Garufi, Gateley, Gaudio,
  Gayathri, Gemme, Genin, Gennai, George, George, Gergely, Ghonge, Ghosh,
  Ghosh, Ghosh, Giacomazzo, Giaime, Giardina, Gibson, Gier, Gill, Glanzer,
  Gniesmer, Godwin, Goetz, Goetz, Gohlke, Goncharov, González, Gopakumar,
  Gossan, Gosselin, Gouaty, Grace, Grado, Granata, Grant, Gras, Grassia, Gray,
  Gray, Greco, Green, Green, Gretarsson, Griggs, Grignani, Grimaldi, Grimm,
  Grote, Grunewald, Gruning, Guidi, Guimaraes, Guixé, Gulati, Guo, Gupta,
  Gupta, Gupta, Gustafson, Gustafson, Haegel, Halim, Hall, Hamilton, Hammond,
  Haney, Hanke, Hanks, Hanna, Hannam, Hannuksela, Hansen, Hanson, Harder,
  Hardwick, Haris, Harms, Harry, Harry, Hasskew, Haster, Haughian, Hayes,
  Healy, Heidmann, Heintze, Heinze, Heitmann, Hellman, Hello, Hemming, Hendry,
  Heng, Hennes, Hennig, Heurs, Hild, Hinderer, Hoback, Hochheim, Hofgard,
  Hofman, Holgado, Holland, Holt, Holz, Hopkins, Horst, Hough, Howell, Hoy,
  Huang, Hübner, Huerta, Huet, Hughey, Hui, Husa, Huttner, Huxford,
  Huynh-Dinh, Idzkowski, Iess, Inchauspe, Ingram, Intini, Isac, Isi, Iyer,
  Jacqmin, Jadhav, Jadhav, James, Jani, Janthalur, Jaranowski, Jariwala, Jaume,
  Jenkins, Jiang, Johns, Jones, Jones, Jones, Jones, Jones, Jonker, Ju, Junker,
  Kalaghatgi, Kalogera, Kamai, Kandhasamy, Kang, Kanner, Kapadia, Karki,
  Kashyap, Kasprzack, Kastaun, Katsanevas, Katsavounidis, Katzman, Kaufer,
  Kawabe, Kéfélian, Keitel, Keivani, Kennedy, Key, Khadka, Khalili, Khan,
  Khan, Khan, Khazanov, Khetan, Khursheed, Kijbunchoo, Kim, Kim, Kim, Kim, Kim,
  Kim, Kim, Kimball, King, Kinley-Hanlon, Kirchhoff, Kissel, Kleybolte,
  Klimenko, Knowles, Koch, Koehlenbeck, Koekoek, Koley, Kondrashov, Kontos,
  Koper, Korobko, Korth, Kovalam, Kozak, Kringel, Krishnendu, Królak,
  Krupinski, Kuehn, Kumar, Kumar, Kumar, Kumar, Kumar, Kuo, Kutynia, Lackey,
  Laghi, Lalande, Lam, Lamberts, Landry, Lane, Lang, Lange, Lantz, Lanza, {La
  Rosa}, Lartaux-Vollard, Lasky, Laxen, Lazzarini, Lazzaro, Leaci, Leavey,
  Lecoeuche, Lee, Lee, Lee, Lee, Lee, Lehmann, Leroy, Letendre, Levin, Li, Li,
  li, Li, Li, Linde, Linker, Linley, Littenberg, Liu, Liu, Llorens-Monteagudo,
  Lo, Lockwood, London, Longo, Lorenzini, Loriette, Lormand, Losurdo, Lough,
  Lousto, Lovelace, Lück, Lumaca, Lundgren, Ma, Macas, Macfoy, MacInnis,
  Macleod, MacMillan, Macquet, Hernandez, Magaña-Sandoval, Magee, Majorana,
  Maksimovic, Malik, Man, Mandic, Mangano, Mansell, Manske, Mantovani, Mapelli,
  Marchesoni, Marion, Márka, Márka, Markakis, Markosyan, Markowitz, Maros,
  Marquina, Marsat, Martelli, Martin, Martin, Martinez, Martynov, Masalehdan,
  Mason, Massera, Masserot, Massinger, Masso-Reid, Mastrogiovanni, Matas,
  Matichard, Mavalvala, Maynard, McCann, McCarthy, McClelland, McCormick,
  McCuller, McGuire, McIsaac, McIver, McManus, McRae, McWilliams, Meacher,
  Meadors, Mehmet, Mehta, Villa, Melatos, Mendell, Mercer, Mereni, Merfeld,
  Merilh, Merritt, Merzougui, Meshkov, Messenger, Messick, Metzdorff, Meyers,
  Meylahn, Mhaske, Miani, Miao, Michaloliakos, Michel, Middleton, Milano,
  Miller, Millhouse, Mills, Milotti, Milovich-Goff, Minazzoli, Minenkov,
  Mishkin, Mishra, Mistry, Mitra, Mitrofanov, Mitselmakher, Mittleman, Mo,
  Mogushi, Mohapatra, Mohite, Molina-Ruiz, Mondin, Montani, Moore, Moraru,
  Morawski, Moreno, Morisaki, Mours, Mow-Lowry, Mozzon, Muciaccia, Mukherjee,
  Mukherjee, Mukherjee, Mukherjee, Mukund, Mullavey, Munch, Muñiz, Murray,
  Nagar, Nardecchia, Naticchioni, Nayak, Neil, Neilson, Nelemans, Nelson, Nery,
  Neunzert, Ng, Ng, Nguyen, Nguyen, Nichols, Nichols, Nissanke, Nocera, Noh,
  North, Nothard, Nuttall, Oberling, O’Brien, Oganesyan, Ogin, Oh, Oh, Ohme,
  Ohta, Okada, Oliver, Olivetto, Oppermann, Oram, O’Reilly, Ormiston, Ortega,
  O’Shaughnessy, Ossokine, Osthelder, Ottaway, Overmier, Owen, Pace, Pagano,
  Page, Pagliaroli, Pai, Pai, Palamos, Palashov, Palomba, Pan, Panda, Pang,
  Pankow, Pannarale, Pant, Paoletti, Paoli, Parida, Parker, Pascucci,
  Pasqualetti, Passaquieti, Passuello, Patricelli, Payne, Pearlstone, Pechsiri,
  Pedersen, Pedraza, Pele, Penn, Perego, Perez, Périgois, Perreca, Perriès,
  Petermann, Pfeiffer, Phelps, Phukon, Piccinni, Pichot, Piendibene,
  Piergiovanni, Pierro, Pillant, Pinard, Pinto, Piotrzkowski, Pirello, Pitkin,
  Plastino, Poggiani, Pong, Ponrathnam, Popolizio, Porter, Powell, Prajapati,
  Prasai, Prasanna, Pratten, Prestegard, Principe, Prodi, Prokhorov, Punturo,
  Puppo, Pürrer, Qi, Quetschke, Quinonez, Raab, Raaijmakers, Radkins,
  Radulesco, Raffai, Rafferty, Raja, Rajan, Rajbhandari, Rakhmanov, Ramirez,
  Ramos-Buades, Rana, Rao, Rapagnani, Raymond, Razzano, Read, Regimbau, Rei,
  Reid, Reitze, Rettegno, Ricci, Richardson, Richardson, Ricker,
  Riemenschneider, Riles, Rizzo, Robertson, Robinet, Rocchi, Rodriguez-Soto,
  Rolland, Rollins, Roma, Romanelli, Romano, Romel, Romero-Shaw, Romie, Rose,
  Rose, Rose, Rosińska, Rosofsky, Ross, Rowan, Rowlinson, Roy, Roy, Roy,
  Ruggi, Rutins, Ryan, Sachdev, Sadecki, Sakellariadou, Salafia, Salconi,
  Saleem, Samajdar, Sanchez, Sanchez, Sanchis-Gual, Sanders, Santiago, Santos,
  Sarin, Sassolas, Sathyaprakash, Sauter, Savage, Savant, Sawant, Sayah,
  Schaetzl, Schale, Scheel, Scheuer, Schmidt, Schnabel, Schofield, Schönbeck,
  Schreiber, Schulte, Schutz, Schwarm, Schwartz, Scott, Scott, Seidel, Sellers,
  Sengupta, Sennett, Sentenac, Sequino, Sergeev, Setyawati, Shaddock, Shaffer,
  Shahriar, Sharma, Sharma, Shawhan, Shen, Shikauchi, Shink, Shoemaker,
  Shoemaker, Shukla, ShyamSundar, Siellez, Sieniawska, Sigg, Singer, Singh,
  Singh, Singha, Singhal, Sintes, Sipala, Skliris, Slagmolen, Slaven-Blair,
  Smetana, Smith, Smith, Somala, Son, Soni, Sorazu, Sordini, Sorrentino,
  Souradeep, Sowell, Spencer, Spera, Srivastava, Srivastava, Staats, Stachie,
  Standke, Steer, Steinke, Steinlechner, Steinlechner, Steinmeyer, Stocks,
  Stops, Stover, Strain, Stratta, Strunk, Sturani, Stuver, Sudhagar, Sudhir,
  Summerscales, Sun, Sunil, Sur, Suresh, Sutton, Swinkels, Szczepańczyk,
  Tacca, Tait, Talbot, Tanasijczuk, Tanner, Tao, Tápai, Tapia, Martin, Tasson,
  Taylor, Tenorio, Terkowski, Thirugnanasambandam, Thomas, Thomas, Thompson,
  Thondapu, Thorne, Thrane, Tinsman, Saravanan, Tiwari, Tiwari, Tiwari, Toland,
  Tonelli, Tornasi, Torres-Forné, Torrie, {Tosta e Melo}, Töyrä, Trail,
  Travasso, Traylor, Tringali, Tripathee, Trovato, Trudeau, Tsang, Tse, Tso,
  Tsukada, Tsuna, Tsutsui, Turconi, Ubhi, Ueno, Ugolini, Unnikrishnan, Urban,
  Usman, Utina, Vahlbruch, Vajente, Valdes, Valentini, Vallisneri, {van Bakel},
  {van Beuzekom}, {van den Brand}, {Van Den Broeck}, Vander-Hyde, {van der
  Schaaf}, {Van Heijningen}, {van Veggel}, Vardaro, Varma, Vass, Vasúth,
  Vecchio, Vedovato, Veitch, Veitch, Venkateswara, Venugopalan, Verkindt,
  Veske, Vetrano, Viceré, Viets, Vinciguerra, Vine, Vinet, Vitale, Vivanco,
  Vo, Vocca, Vorvick, Vyatchanin, Wade, Wade, Wade, Walet, Walker, Wallace,
  Wallace, Walsh, Wang, Wang, Wang, Wang, Ward, Warden, Warner, Was, Watchi,
  Weaver, Wei, Weinert, Weinstein, Weiss, Wellmann, Wen, Weßels, Westhouse,
  Wette, Whelan, Whiting, Whittle, Wilken, Williams, Williams, Williamson,
  Willis, Willke, Winkler, Wipf, Wittel, Woan, Woehler, Wofford, Wong, Wright,
  Wu, Wysocki, Xiao, Yamamoto, Yang, Yang, Yang, Yap, Yazback, Yeeles, Yu, Yu,
  Yuen, Zadrożny, Zadrożny, Zanolin, Zelenova, Zendri, Zevin, Zhang, Zhang,
  Zhang, Zhao, Zhao, Zhou, Zhou, Zhu, Zimmerman, Zucker, \&
  Zweizig}]{Abbott:2019ebz}
Abbott, R., Abbott, T., Abraham, S., {et~al.} 2021, SoftwareX, 13, 100658,
  \dodoi{https://doi.org/10.1016/j.softx.2021.100658}

\bibitem[{Acernese {et~al.}(2015)}]{TheVirgo:2014hva}
Acernese, F., {et~al.} 2015, Class. Quantum Grav., 32, 024001,
  \dodoi{10.1088/0264-9381/32/2/024001}

\bibitem[{Ali-Haïmoud {et~al.}(2017)Ali-Haïmoud, Kovetz, \&
  Kamionkowski}]{Ali-Haimoud:2017rtz}
Ali-Haïmoud, Y., Kovetz, E.~D., \& Kamionkowski, M. 2017, Phys. Rev. D, 96,
  123523, \dodoi{10.1103/PhysRevD.96.123523}

\bibitem[{Belczynski {et~al.}(2016)}]{Belczynski:2016jno}
Belczynski, K., {et~al.} 2016, Astron. Astrophys., 594, A97,
  \dodoi{10.1051/0004-6361/201628980}

\bibitem[{{Bird} {et~al.}(2016){Bird}, {Cholis}, {Mu{\~n}oz},
  {Ali-Ha{\"\i}moud}, {Kamionkowski}, {Kovetz}, {Raccanelli}, \&
  {Riess}}]{Bird:2016}
{Bird}, S., {Cholis}, I., {Mu{\~n}oz}, J.~B., {et~al.} 2016, \prl, 116, 201301,
  \dodoi{10.1103/PhysRevLett.116.201301}

\bibitem[{Biswas {et~al.}(2009)Biswas, Brady, Creighton, \&
  Fairhurst}]{Biswas:2007ni}
Biswas, R., Brady, P.~R., Creighton, J. D.~E., \& Fairhurst, S. 2009, Class.
  Quantum Grav., 26, 175009, \dodoi{10.1088/0264-9381/26/17/175009}

\bibitem[{Brown(2004)}]{findchirp}
Brown, D.~A. 2004, PhD thesis, University of Wisconsin--Milwaukee

\bibitem[{Brown {et~al.}(2012)Brown, Lundgren, \& O'Shaughnessy}]{Brown:2012gs}
Brown, D.~A., Lundgren, A., \& O'Shaughnessy, R. 2012, Phys. Rev. D, 86,
  064020, \dodoi{10.1103/PhysRevD.86.064020}

\bibitem[{Byrnes {et~al.}(2018)Byrnes, Hindmarsh, Young, \&
  Hawkins}]{Byrnes:2018clq}
Byrnes, C.~T., Hindmarsh, M., Young, S., \& Hawkins, M. R.~S. 2018, JCAP, 08,
  041, \dodoi{10.1088/1475-7516/2018/08/041}

\bibitem[{Calder\'on~Bustillo {et~al.}(2020)Calder\'on~Bustillo, Sanchis-Gual,
  Torres-Forn\'e, \& Font}]{CalderonBustillo:2020odh}
Calder\'on~Bustillo, J., Sanchis-Gual, N., Torres-Forn\'e, A., \& Font, J.~A.
  2020.
\newblock \doarXiv{2009.01066}

\bibitem[{Chen \& Huang(2018)}]{Chen:2018czv}
Chen, Z.-C., \& Huang, Q.-G. 2018, Astrophys. J., 864, 61,
  \dodoi{10.3847/1538-4357/aad6e2}

\bibitem[{Chiba \& Yokoyama(2017)}]{Chiba:2017rvs}
Chiba, T., \& Yokoyama, S. 2017, PTEP, 2017, 083E01,
  \dodoi{10.1093/ptep/ptx087}

\bibitem[{{Cholis} {et~al.}(2016){Cholis}, {Kovetz}, {Ali-Ha{\"\i}moud},
  {Bird}, {Kamionkowski}, {Mu{\~n}oz}, \& {Raccanelli}}]{Cholis:2016}
{Cholis}, I., {Kovetz}, E.~D., {Ali-Ha{\"\i}moud}, Y., {et~al.} 2016, \prd, 94,
  084013, \dodoi{10.1103/PhysRevD.94.084013}

\bibitem[{Clesse \& Garc\'\i{}a-Bellido(2017)}]{Clesse:2016vqa}
Clesse, S., \& Garc\'\i{}a-Bellido, J. 2017, Phys. Dark Univ., 15, 142,
  \dodoi{10.1016/j.dark.2016.10.002}

\bibitem[{De~Luca {et~al.}(2019)De~Luca, Desjacques, Franciolini, Malhotra, \&
  Riotto}]{DeLuca:2019buf}
De~Luca, V., Desjacques, V., Franciolini, G., Malhotra, A., \& Riotto, A. 2019,
  JCAP, 05, 018, \dodoi{10.1088/1475-7516/2019/05/018}

\bibitem[{De~Luca {et~al.}(2020{\natexlab{a}})De~Luca, Franciolini, Pani, \&
  Riotto}]{DeLuca:2020qqa}
De~Luca, V., Franciolini, G., Pani, P., \& Riotto, A. 2020{\natexlab{a}}, JCAP,
  06, 044, \dodoi{10.1088/1475-7516/2020/06/044}

\bibitem[{De~Luca {et~al.}(2020{\natexlab{b}})De~Luca, Franciolini, Pani, \&
  Riotto}]{DeLuca:2020bjf}
---. 2020{\natexlab{b}}, JCAP, 04, 052, \dodoi{10.1088/1475-7516/2020/04/052}

\bibitem[{Fishbach \& Holz(2020)}]{Fishbach:2020qag}
Fishbach, M., \& Holz, D.~E. 2020, Astrophys. J. Lett., 904, L26,
  \dodoi{10.3847/2041-8213/abc827}

\bibitem[{Gayathri {et~al.}(2020)Gayathri, Healy, Lange, O'Brien, Szczepanczyk,
  Bartos, Campanelli, Klimenko, Lousto, \& O'Shaughnessy}]{Gayathri:2020coq}
Gayathri, V., Healy, J., Lange, J., {et~al.} 2020.
\newblock \doarXiv{2009.05461}

\bibitem[{Georg \& Watson(2017)}]{Georg:2017mqk}
Georg, J., \& Watson, S. 2017, JHEP, 09, 138, \dodoi{10.1007/JHEP09(2017)138}

\bibitem[{Green \& Kavanagh(2020)}]{Green:2020jor}
Green, A.~M., \& Kavanagh, B.~J. 2020.
\newblock \doarXiv{2007.10722}

\bibitem[{Harry {et~al.}(2009)Harry, Allen, \& Sathyaprakash}]{Harry:2009ea}
Harry, I.~W., Allen, B., \& Sathyaprakash, B. 2009, Phys. Rev. D, 80, 104014,
  \dodoi{10.1103/PhysRevD.80.104014}

\bibitem[{Jedamzik(1997)}]{Jedamzik:1996mr}
Jedamzik, K. 1997, Phys. Rev. D, 55, 5871, \dodoi{10.1103/PhysRevD.55.R5871}

\bibitem[{{Kagra Collaboration} {et~al.}(2019){Kagra Collaboration}, {Akutsu},
  {Ando}, {Arai}, {Arai}, {Araki}, {Araya}, {Aritomi}, {Asada}, {Aso},
  {Atsuta}, {Awai}, {Bae}, {Baiotti}, {Barton}, {Cannon}, {Capocasa}, {Chen},
  {Chiu}, {Cho}, {Chu}, {Craig}, {Creus}, {Doi}, {Eda}, {Enomoto}, {Flaminio},
  {Fujii}, {Fujimoto}, {Fukunaga}, {Fukushima}, {Furuhata}, {Haino},
  {Hasegawa}, {Hashino}, {Hayama}, {Hirobayashi}, {Hirose}, {Hsieh}, {Huang},
  {Ikenoue}, {Inoue}, {Ioka}, {Itoh}, {Izumi}, {Kaji}, {Kajita}, {Kakizaki},
  {Kamiizumi}, {Kanbara}, {Kanda}, {Kanemura}, {Kaneyama}, {Kang}, {Kasuya},
  {Kataoka}, {Kawai}, {Kawamura}, {Kawasaki}, {Kim}, {Kim}, {Kim}, {Kim},
  {Kim}, {Kimura}, {Kinugawa}, {Kirii}, {Kitaoka}, {Kitazawa}, {Kojima},
  {Kokeyama}, {Komori}, {Kong}, {Kotake}, {Kozu}, {Kumar}, {Kuo}, {Kuroyanagi},
  {Lee}, {Lee}, {Lee}, {Leonardi}, {Lin}, {Lin}, {Liu}, {Liu}, {Majorana},
  {Mano}, {Marchio}, {Matsui}, {Matsushima}, {Michimura}, {Mio}, {Miyakawa},
  {Miyamoto}, {Miyamoto}, {Miyo}, {Miyoki}, {Morii}, {Morisaki}, {Moriwaki},
  {Morozumi}, {Musha}, {Nagano}, {Nagano}, {Nakamura}, {Nakamura}, {Nakano},
  {Nakano}, {Nakao}, {Narikawa}, {Naticchioni}, {Nguyen Quynh}, {Ni},
  {Nishizawa}, {Obuchi}, {Ochi}, {Oh}, {Oh}, {Ohashi}, {Ohishi}, {Ohkawa},
  {Okutomi}, {Ono}, {Oohara}, {Ooi}, {Pan}, {Park}, {Pe{\~n}a Arellano},
  {Pinto}, {Sago}, {Saijo}, {Saitou}, {Saito}, {Sakai}, {Sakai}, {Sakai},
  {Sasai}, {Sasaki}, {Sasaki}, {Sato}, {Sato}, {Sato}, {Sekiguchi}, {Seto},
  {Shibata}, {Shimoda}, {Shinkai}, {Shishido}, {Shoda}, {Somiya}, {Son},
  {Suemasa}, {Suzuki}, {Suzuki}, {Tagoshi}, {Tahara}, {Takahashi}, {Takahashi},
  {Takamori}, {Takeda}, {Tanaka}, {Tanaka}, {Tanaka}, {Tanioka}, {Tapia San
  Martin}, {Tatsumi}, {Tomaru}, {Tomura}, {Travasso}, {Tsubono}, {Tsuchida},
  {Uchikata}, {Uchiyama}, {Uehara}, {Ueki}, {Ueno}, {Uraguchi}, {Ushiba}, {van
  Putten}, {Vocca}, {Wada}, {Wakamatsu}, {Watanabe}, {Xu}, {Yamada},
  {Yamamoto}, {Yamamoto}, {Yamamoto}, {Yamamoto}, {Yamamoto}, {Yokogawa},
  {Yokoyama}, {Yokozawa}, {Yoon}, {Yoshioka}, {Yuzurihara}, {Zeidler}, \&
  {Zhu}}]{KAGRA}
{Kagra Collaboration}, {Akutsu}, T., {Ando}, M., {et~al.} 2019, Nature
  Astronomy, 3, 35, \dodoi{10.1038/s41550-018-0658-y}

\bibitem[{{Kritos} {et~al.}(2020){Kritos}, {De Luca}, {Franciolini},
  {Kehagias}, \& {Riotto}}]{2012.03585}
{Kritos}, K., {De Luca}, V., {Franciolini}, G., {Kehagias}, A., \& {Riotto}, A.
  2020, arXiv e-prints, arXiv:2012.03585.
\newblock \doarXiv{2012.03585}

\bibitem[{Marchant {et~al.}(2019)Marchant, Renzo, Farmer, Pappas, Taam,
  de~Mink, \& Kalogera}]{Marchant:2018kun}
Marchant, P., Renzo, M., Farmer, R., {et~al.} 2019, Astrophys. J., 882, 36,
  \dodoi{10.3847/1538-4357/ab3426}

\bibitem[{Mirbabayi {et~al.}(2020)Mirbabayi, Gruzinov, \&
  Noreña}]{Mirbabayi:2019uph}
Mirbabayi, M., Gruzinov, A., \& Noreña, J. 2020, JCAP, 03, 017,
  \dodoi{10.1088/1475-7516/2020/03/017}

\bibitem[{{Moore} {et~al.}(2018){Moore}, {Robson}, {Loutrel}, \&
  {Yunes}}]{Moore:2018}
{Moore}, B., {Robson}, T., {Loutrel}, N., \& {Yunes}, N. 2018, Classical and
  Quantum Gravity, 35, 235006, \dodoi{10.1088/1361-6382/aaea00}

\bibitem[{{Moore} \& {Yunes}(2019{\natexlab{a}})}]{Moore:2019a}
{Moore}, B., \& {Yunes}, N. 2019{\natexlab{a}}, Classical and Quantum Gravity,
  36, 185003, \dodoi{10.1088/1361-6382/ab3778}

\bibitem[{{Moore} \& {Yunes}(2019{\natexlab{b}})}]{Moore:2019b}
---. 2019{\natexlab{b}}, arXiv e-prints, arXiv:1910.01680.
\newblock \doarXiv{1910.01680}

\bibitem[{Nakamura {et~al.}(1997)Nakamura, Sasaki, Tanaka, \&
  Thorne}]{Nakamura:1997sm}
Nakamura, T., Sasaki, M., Tanaka, T., \& Thorne, K.~S. 1997, \apj, 487, L139

\bibitem[{{Nishikawa} {et~al.}(2019){Nishikawa}, {Kovetz}, {Kamionkowski}, \&
  {Silk}}]{10.1103/PhysRevD.99.043533}
{Nishikawa}, H., {Kovetz}, E.~D., {Kamionkowski}, M., \& {Silk}, J. 2019, \prd,
  99, 043533, \dodoi{10.1103/PhysRevD.99.043533}

\bibitem[{Nitz \& Capano(2021)}]{Nitz:2020mga}
Nitz, A.~H., \& Capano, C.~D. 2021, Astrophys. J. Lett., 907, L9,
  \dodoi{10.3847/2041-8213/abccc5}

\bibitem[{Nitz {et~al.}(2019{\natexlab{a}})Nitz, Lenon, \&
  Brown}]{Nitz:2019spj}
Nitz, A.~H., Lenon, A., \& Brown, D.~A. 2019{\natexlab{a}}, Astrophys. J., 890,
  1, \dodoi{10.3847/1538-4357/ab6611}

\bibitem[{Nitz {et~al.}(2013)Nitz, Lundgren, Brown, Ochsner, Keppel,
  {et~al.}}]{Nitz:2013mxa}
Nitz, A.~H., Lundgren, A., Brown, D.~A., {et~al.} 2013, Phys.~Rev.~D, 88,
  124039, \dodoi{10.1103/PhysRevD.88.124039}

\bibitem[{Nitz \& Wang(2021)}]{Nitz:2020bdb}
Nitz, A.~H., \& Wang, Y.-F. 2021, Phys. Rev. Lett., 126, 021103,
  \dodoi{10.1103/PhysRevLett.126.021103}

\bibitem[{Nitz {et~al.}(2018)Nitz, Harry, Willis, Biwer, Brown, Pekowsky,
  Dal~Canton, Williamson, Dent, Capano, Massinger, Lenon, Nielsen, \&
  Cabero}]{pycbc-github}
Nitz, A.~H., Harry, I.~W., Willis, J.~L., {et~al.} 2018, {PyCBC Software},
  \url{https://github.com/gwastro/pycbc},  GitHub

\bibitem[{Nitz {et~al.}(2019{\natexlab{b}})Nitz, Dent, Davies, Kumar, Capano,
  Harry, Mozzon, Nuttall, Lundgren, \& Tápai}]{Nitz:2019hdf}
Nitz, A.~H., Dent, T., Davies, G.~S., {et~al.} 2019{\natexlab{b}}, Astrophys.
  J., 891, 123, \dodoi{10.3847/1538-4357/ab733f}

\bibitem[{Postnov {et~al.}(2019)Postnov, Kuranov, \&
  Mitichkin}]{Postnov:2019pkd}
Postnov, K., Kuranov, A., \& Mitichkin, N. 2019, Phys. Usp., 62, 1153,
  \dodoi{10.3367/UFNe.2019.04.038593}

\bibitem[{Punturo {et~al.}(2010)Punturo, Abernathy, Acernese, Allen, Andersson,
  Arun, Barone, Barr, Barsuglia, Beker, Beveridge, Birindelli, Bose, Bosi,
  Braccini, Bradaschia, Bulik, Calloni, Cella, Mottin, Chelkowski, Chincarini,
  Clark, Coccia, Colacino, Colas, Cumming, Cunningham, Cuoco, Danilishin,
  Danzmann, Luca, Salvo, Dent, Rosa, Fiore, Virgilio, Doets, Fafone, Falferi,
  Flaminio, Franc, Frasconi, Freise, Fulda, Gair, Gemme, Gennai, Giazotto,
  Glampedakis, Granata, Grote, Guidi, Hammond, Hannam, Harms, Heinert, Hendry,
  Heng, Hennes, Hild, Hough, Husa, Huttner, Jones, Khalili, Kokeyama, Kokkotas,
  Krishnan, Lorenzini, Lück, Majorana, Mandel, Mandic, Martin, Michel,
  Minenkov, Morgado, Mosca, Mours, Müller{\textendash}Ebhardt, Murray,
  Nawrodt, Nelson, Oshaughnessy, Ott, Palomba, Paoli, Parguez, Pasqualetti,
  Passaquieti, Passuello, Pinard, Poggiani, Popolizio, Prato, Puppo, Rabeling,
  Rapagnani, Read, Regimbau, Rehbein, Reid, Rezzolla, Ricci, Richard, Rocchi,
  Rowan, Rüdiger, Sassolas, Sathyaprakash, Schnabel, Schwarz, Seidel, Sintes,
  Somiya, Speirits, Strain, Strigin, Sutton, Tarabrin, Thüring, van~den Brand,
  van Leewen, van Veggel, van~den Broeck, Vecchio, Veitch, Vetrano, Vicere,
  Vyatchanin, Willke, Woan, Wolfango, \& Yamamoto}]{et}
Punturo, M., Abernathy, M., Acernese, F., {et~al.} 2010, Classical and Quantum
  Gravity, 27, 194002, \dodoi{10.1088/0264-9381/27/19/194002}

\bibitem[{Reitze {et~al.}(2019)}]{cewhitepaper}
Reitze, D., {et~al.} 2019, Bull. Am. Astron. Soc., 51, 035.
\newblock \doarXiv{1907.04833}

\bibitem[{Romero-Shaw {et~al.}(2020)Romero-Shaw, Lasky, Thrane, \&
  Bustillo}]{Romero-Shaw:2020thy}
Romero-Shaw, I.~M., Lasky, P.~D., Thrane, E., \& Bustillo, J.~C. 2020,
  Astrophys. J. Lett., 903, L5, \dodoi{10.3847/2041-8213/abbe26}

\bibitem[{Roulet {et~al.}(2020)Roulet, Venumadhav, Zackay, Dai, \&
  Zaldarriaga}]{Roulet:2020wyq}
Roulet, J., Venumadhav, T., Zackay, B., Dai, L., \& Zaldarriaga, M. 2020, Phys.
  Rev. D, 102, 123022, \dodoi{10.1103/PhysRevD.102.123022}

\bibitem[{Sasaki {et~al.}(2016)Sasaki, Suyama, Tanaka, \&
  Yokoyama}]{Sasaki:2016jop}
Sasaki, M., Suyama, T., Tanaka, T., \& Yokoyama, S. 2016, Phys. Rev. Lett.,
  117, 061101, \dodoi{10.1103/PhysRevLett.117.061101}

\bibitem[{Stevenson {et~al.}(2019)Stevenson, Sampson, Powell, Vigna-Gómez,
  Neijssel, Szécsi, \& Mandel}]{Stevenson:2019rcw}
Stevenson, S., Sampson, M., Powell, J., {et~al.} 2019,
  \dodoi{10.3847/1538-4357/ab3981}

\bibitem[{{Suwa} {et~al.}(2018){Suwa}, {Yoshida}, {Shibata}, {Umeda}, \&
  {Takahashi}}]{minmass2}
{Suwa}, Y., {Yoshida}, T., {Shibata}, M., {Umeda}, H., \& {Takahashi}, K. 2018,
  \mnras, 481, 3305, \dodoi{10.1093/mnras/sty2460}

\bibitem[{{Timmes} {et~al.}(1996){Timmes}, {Woosley}, \& {Weaver}}]{minmass1}
{Timmes}, F.~X., {Woosley}, S.~E., \& {Weaver}, T.~A. 1996, \apj, 457, 834,
  \dodoi{10.1086/176778}

\bibitem[{Usman {et~al.}(2016)}]{Usman:2015kfa}
Usman, S.~A., {et~al.} 2016, Class. Quant. Grav., 33, 215004,
  \dodoi{10.1088/0264-9381/33/21/215004}

\bibitem[{Vallisneri {et~al.}(2015)Vallisneri, Kanner, Williams, Weinstein, \&
  Stephens}]{Vallisneri:2014vxa}
Vallisneri, M., Kanner, J., Williams, R., Weinstein, A., \& Stephens, B. 2015,
  J. Phys. Conf. Ser., 610, 012021, \dodoi{10.1088/1742-6596/610/1/012021}

\bibitem[{Wang \& Nitz(2021)}]{Wang:2021qsu}
Wang, Y.-F., \& Nitz, A.~H. 2021.
\newblock \doarXiv{2101.12269}

\bibitem[{Widerin \& Schmid(1998)}]{Widerin:1998my}
Widerin, P., \& Schmid, C. 1998.
\newblock \doarXiv{astro-ph/9808142}

\bibitem[{Woosley(2017)}]{Woosley:2016hmi}
Woosley, S.~E. 2017, Astrophys. J., 836, 244,
  \dodoi{10.3847/1538-4357/836/2/244}

\bibitem[{Woosley(2019)}]{Woosley_2019}
---. 2019, The Astrophysical Journal, 878, 49, \dodoi{10.3847/1538-4357/ab1b41}

\end{thebibliography}

\end{CJK*}
\end{document}